\documentclass[preprint,amsmath,amssymb,superscriptaddress]{revtex4-1}
\usepackage{graphicx}
\usepackage[caption=false]{subfig}
\usepackage{dcolumn}
\usepackage{bm}
\usepackage{ulem}
\usepackage[utf8]{inputenc}
\usepackage[T1]{fontenc}
\setcitestyle{square}
\usepackage{color}
\usepackage{xcolor}

\usepackage{hyperref}       
\usepackage{url}            
\usepackage{booktabs}       
\usepackage{amsfonts}       
\usepackage{nicefrac}       
\usepackage{lipsum}
\usepackage{wasysym}
\usepackage{lineno}

\usepackage{natbib}
\usepackage{float}

\begin{document}

\title{Experimental characterisation of a single-shot spectrometer for high-flux, GeV-scale gamma-ray beams}

\newcommand{\JAI}{The John Adams Institute for Accelerator Science, Imperial College London, London, SW7 2AZ, UK}

\newcommand{\GOLP}{GoLP/Instituto de Plasmas e Fus\~{a}o Nuclear, Instituto Superior T\'{e}cnico, U.L., Lisboa 1049-001, Portugal}

\newcommand{\CLF}{Central Laser Facility, STFC Rutherford Appleton Laboratory, Didcot OX11 0QX, UK}

\newcommand{\UCL}{Department of Physics and Astronomy, University College London, London WC1E 6BT, UK}

\newcommand{\LMU}{Fakult\"at f\"ur Physik, Ludwig-Maximilians-Universit\"at M\"unchen, D-85748 Garching, Germany}
\newcommand{\MPQ}{Max-Planck-Institut f\"ur Quantenoptik, Hans-Kopfermann-Str. 1, D-85748 Garching, Germany}

\newcommand{\DESY}{Deutsches Elektronen-Synchrotron DESY, Notkestr. 85, 22607 Hamburg, Germany}

\newcommand{\CI}{The Cockcroft Institute, Keckwick Lane, Daresbury, WA4 4AD, United Kingdom}

\newcommand{\LANCS}{Physics Department, Lancaster University, Lancaster LA1 4YB, United Kingdom}

\newcommand{\UMICH}{Center for Ultrafast Optical Science, University of Michigan, Ann Arbor, MI 48109-2099, USA}

\newcommand{\SUPA}{SUPA, Department of Physics, University of Strathclyde, Glasgow G4 0NG, UK}

\newcommand{\LLNL}{Lawrence Livermore National Laboratory (LLNL), P.O. Box 808, Livermore, California 94550, USA}

\newcommand{\DLS}{Diamond Light Source, Harwell Science and Innovation Campus, Fermi Avenue, Didcot OX11 0DE, UK}

\newcommand{\YORK}{York Plasma Institute, Department of Physics, University of York, York YO10 5DD, UK}

\newcommand{\ELI}{ELI-Beamlines, Institute of Physics, Academy of Sciences of the Czech Republic, 18221 Prague, Czech Republic}

\newcommand{\LUND}{Department of Physics, Lund University, P.O. Box 118, S-22100, Lund, Sweden}

\newcommand{\QUB}{Centre for Plasma Physics,
  School of Mathematics and Physics,
  Queen's University Belfast
 , BT7 1NN, Belfast United Kingdom}

\newcommand{\LPGP}{LPGP, CNRS, Université Paris-Saclay, 91405 Orsay, France}
\newcommand{\LLR}{LLR, CNRS, École Polytechnique, Palaiseau, France}
\newcommand{\LULI}{LULI—CNRS, CEA, UPMC Université Paris 06: Sorbonne Université, Ecole Polytechnique, Institut Polytechnique de Paris, Palaiseau, France}
\newcommand{\Paris}{Université Paris-Saclay, CEA, CNRS, LIDYL, 91191 Gif-sur-Yvette, France}


\author{N.~Cavanagh}
\affiliation{\QUB}

\author{K.~Fleck}
\affiliation{\QUB}

\author{M.J.V.~Streeter}
\affiliation{\QUB}

\author{E. Gerstmayr}
\affiliation{\QUB}

\author{L.T.~Dickson}
\affiliation{\LPGP}

\author{C.~Ballage}
\affiliation{\LPGP}

\author{R.~Cadas}
\affiliation{\LPGP}

\author{L.~Calvin}
\affiliation{\QUB}

\author{S.~Dobosz~Dufrénoy}
\affiliation{\Paris}

\author{I.~Moulanier}
\affiliation{\LPGP}

\author{L.~Romagnani}
\affiliation{\LULI}

\author{O.~Vasilovici}
\affiliation{\LPGP}

\author{A.~Whitehead}
\affiliation{\Paris}

\author{A.~Specka}
\affiliation{\LLR}

\author{B.~Cros}
\affiliation{\LPGP}

\author{G.Sarri}
\email{g.sarri@qub.ac.uk}
\affiliation{\QUB}

\begin{abstract}
We report on the first experimental characterisation of a gamma-ray spectrometer designed to spectrally resolve high-flux photon beams with energies in the GeV range. The spectrometer has been experimentally characterised using a bremsstrahlung source obtained at the Apollon laser facility during the interaction of laser-wakefield accelerated electron beams (maximum energy of 1.7 GeV and overall charge of 207$\pm$62 pC) with a 1 mm thick tantalum target. Experimental data confirms the possibility of performing single-shot measurements, without the need for accumulation, with a high signal to noise ratio. Scaling the results to photons in the multi-GeV range suggests the possibility of achieving percent-level energy resolution, as required, for instance, by the next generation of experiments in strong-field quantum electrodynamics. 
\end{abstract}


\maketitle

\section{Introduction}\label{introduction}
Gamma-ray beams with energies per photon in the multi-MeV up to the GeV range play a central role in a wide range of physical phenomena and are appealing for potential practical applications of great interest. 
Typically, high-flux sources of high-energy gamma-ray beams can be produced in the laboratory either via bremsstrahlung of an electron beam propagating through a solid target (see, for
instance, Ref. \cite{Schumaker:2014, Glinec:2005, Giulietti:2008}), or via inverse Compton scattering of an electron beam in the field
of an intense laser \cite{Sarri:2014, Yan:2017}. Other mechanisms, exploiting the near-term generation of multi-PW
laser facilities, include direct laser irradiation of solids \cite{Ridgers:2012, Nakamura:2012}, or electromagnetic cascades \cite{Gonoskov:2017}. 
Producing high-energy, well-characterised gamma-ray sources is also essential not only to study nuclear phenomena (see, for instance, Ref. \cite{Nuclear_sources}), but also to provide information of the physics at play in strong-field quantum electrodynamics (SFQED) \cite{Gonoskov_RMP}. 

For example, Compton scattering of an ultra-relativistic electron in the field of an intense laser is strongly affected by quantum effects such as quantum radiation reaction \cite{Cole:2018,Poder:2018}, stochastic photon emission \cite{Blackburn:2014}, and electron-positron pair production \cite{Ilderton:2011, Grismayer:2016}, which are predicted to significantly modify the spectrum and angular distribution of the emitted photons. While experimental work in this area has traditionally focused on characterising the electron and positron populations produced during these interactions (see, e.g. Refs. \cite{Cole:2018, Poder:2018, Kettle:2021,Salgado}), it is now established that key signatures of SFQED phenomena, such as electron mass dressing in the laser field and multi-photon Compton scattering \cite{Sarri:2014}, are embedded also in the spectral and spatial properties of the Compton-scattered gamma-ray beams.
Further, direct photon-photon scattering experiments in the laboratory do require high-energy and high-flux photon sources \cite{Kettle:2021, Hartin:2019, Marklund:2006}; their detailed experimental characterisation is thus an essential requisite to study this elusive phenomenon in the laboratory.

Developing a detector able to precisely measure the spectrum of Compton-scattered or high-energy bremsstrahlung photons is thus crucial in order to advance our understanding of SFQED, justifying the inclusion of gamma-ray spectrometers (based on the design discussed in Ref. \cite{Fleck:2020}) in the design of large-scale SFQED experiments such as the E-320 at SLAC \cite{E320,NaranjoFACET-II} and LUXE at the Eu.XFEL \cite{Abramowicz_2021,LUXE_TDR}.

Spectrally resolving photon beams that are both high-flux and high-energy is a challenging task: to date, gamma ray spectrometers either operate in the multi-GeV regime but at low flux \cite{Schumaker:2014}, or at high flux but low energy (up to tens of MeV) \cite{Corvan:2014,Behm:2018}, or in a narrow spectral region \cite{Barbosa:2015}. 

Here, we provide the first experimental demonstration of the performance of a design recently proposed in Ref. \cite{Fleck:2020}, where the spectrum of high-flux GeV-scale photon beams are obtained, on-shot and non invasively, from the deconvolution of the spectrum of electron and positron pairs produced during the propagation of the gamma-ray beam through a converter target. 
The spectrometer has been tested at the Apollon laser facility \cite{Apollon:2020}, where it was used to spectrally resolve high-flux GeV-scale gamma-ray beams produced via bremsstrahlung during the interaction of a laser-wakefield accelerated electron beam with mm-scale high-Z converter targets. Advancing the first proposal reported in Ref. \cite{Fleck:2020}, we detail here a deconvolution process based on Bayesian inference, which provides regularisation of an otherwise
ill-conditioned problem and allows for an improved treatment of uncertainties in the gamma-ray spectrum.
The experimental results are in excellent agreement with numerical modelling, confirming the feasibility of a detector design of this kind to spectrally resolve high-flux GeV-scale photon beams with high spectral resolution. 

The paper is organised as follows: Section II describes the experimental setup, the typical electron beams obtained from the laser wakefield acceleration, and the expected gamma-ray spectrum after their interaction with a 1 mm tantalum target. Section III shows the resulting signal recorded by the gamma-ray spectrometer. Section IV shows the results of the signal deconvolution to extract the experimental photon spectrum. Finally Section V shows how percent-level accuracy can be achieved at higher photon energies, in regimes of interest to planned SFQED experiments such as E-320 \cite{E320} and LUXE \cite{Abramowicz_2021,LUXE_TDR}. In the appendix, the details of the deconvolution algorithm are given.

\section{Experimental Setup}\label{experimental_setup}
\begin{figure}[h!]
    \centering
    \includegraphics[width=0.8\textwidth]{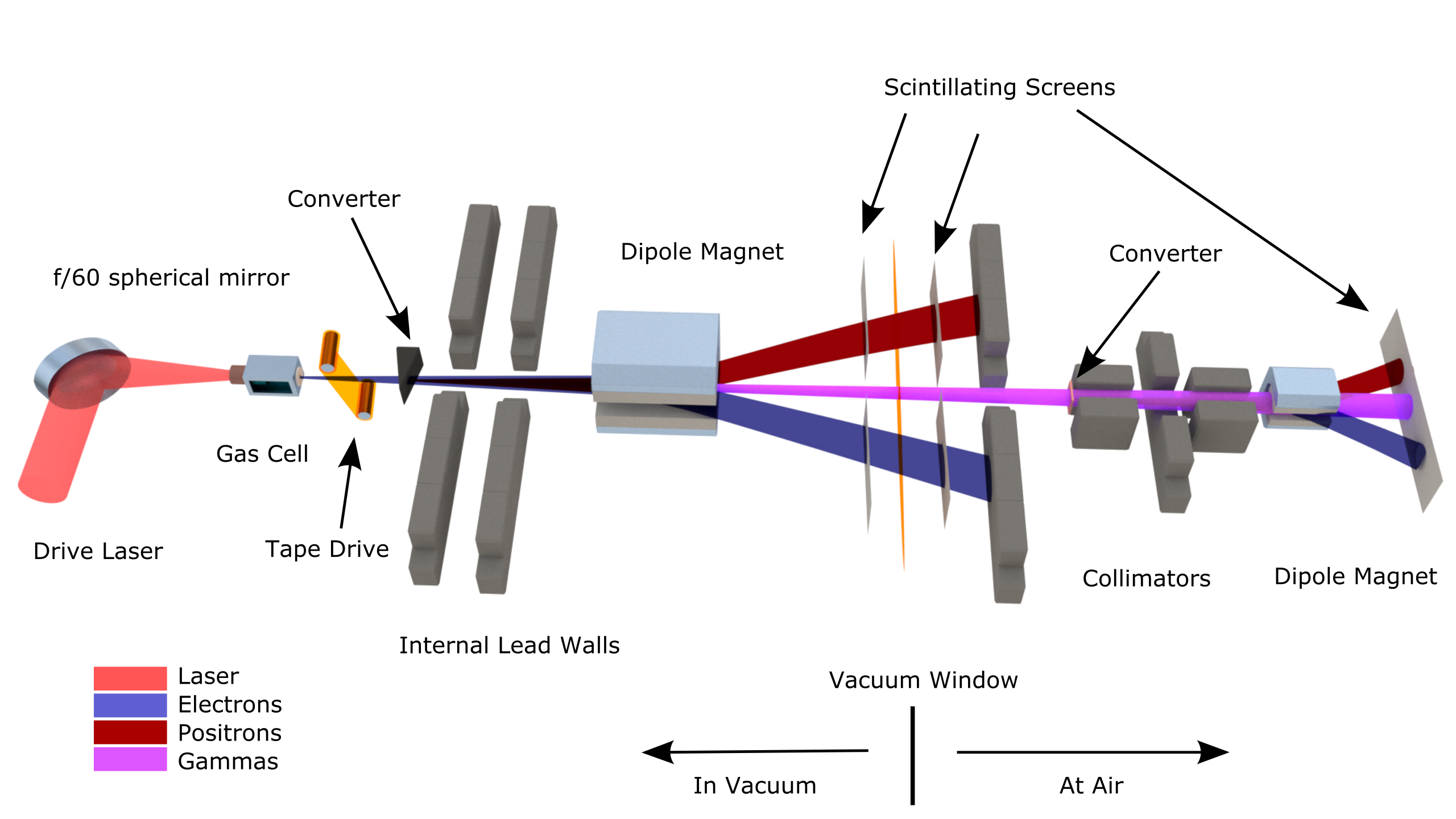}
    \caption{Top-view schematic of the experimental setup.}
    \label{fig:Apollon_layout}
\end{figure}

The experiment (sketched in Fig. \ref{fig:Apollon_layout}) was  performed at the Apollon Laser Facility, which delivered laser pulses with an on-target energy of $12.5\pm 2.5$ J in a $37.5\pm 12.5$ fs FWHM pulse duration at a repetition rate of 1 shot/min.
The laser pulses were focused using a $f$/60 spherical mirror down to a focal spot with a $66\,\mu\textrm{m}$ FWHM containing $37\%$ of the energy, resulting in a peak intensity of $(2.8 \pm 1.1) \times 10^{18}$ Wcm$^{-2}$.
The laser was focused onto a gas cell target with  a variable length between $10$ mm and $25$ mm. This was used to accelerate, via laser-wakefield acceleration (LWFA) \cite{Esarey_RMP}, GeV-scale high-charge electron beams to drive the bremsstrahlung source. The maximum charge and peak energy of the electron beam was consistently obtained for a plasma consisting of 98\% hydrogen and 2\% nitrogen, with an electron density of $10^{18}$ $\mathrm{cm}^{-3}$, which was kept constant for all the data presented hereafter. 
High-energy and high-flux photon beams were produced by directing the electron beams onto a wedged tantalum converter target so that the thickness of the solid traversed by the electron beam could be remotely controlled by translating the converter.
The results presented in this article were obtained using a target with a thickness of 1 mm, placed 69 mm from the rear surface of the gas cell.
The residual laser exiting the gas-cell was removed by
reflection from a self-generated plasma mirror on the surface of a 125 $\mu$m polyimide tape which was replenished after every shot. The tape target was kept for all the experimental data shown here.

Two lead walls, each of $10$ cm thickness with an on-axis bore of $1$ cm diameter, were positioned $14.7$ cm behind the converter, separated by $20$ cm, to provide shielding for the detectors in the vacuum chamber. 
The bore dimension was chosen such that the angular acceptance (full angle) was $12.6$ mrad, sufficient to minimise noise while ensuring unaffected propagation of the gamma-ray beam. After the lead walls, a magnetic dipole with an integrated field strength $0.46$ $\mathrm{T}\cdot\mathrm{m}$ was used to deflect the scattered electrons and generated positrons onto a pair of YAG scintillator screens. 
A $200$ $\mu$m thick Kapton vacuum window with an exterior 4 mm Perspex layer was placed at the rear of the vacuum chamber to allow for the gamma-ray beam to propagate onto the gamma-ray spectrometer, which was placed in air approximately 152 cm from the first converter target. The thicknesses of the Kapton-Perspex layers and the air gap between the chamber window and spectrometer converter are less than $1\%$ of the radiation lengths of the respective materials, hence minimal distortion of the gamma-ray beam is expected as it propagates from source to the spectrometer. This is confirmed by the numerical simulations shown below.

The spectrometer comprised a 225 $\mu$m tantalum foil, which converted a small fraction ($\sim 4\%$) of the radiation into relativistic electron-positron pairs. The generated pairs and remaining gamma-ray radiation then propagated through two collimators, with apertures of $4$ mm and $5$ mm respectively in the dispersion axis, resulting in an acceptance angle of $16$ mrad. 

A 5 cm, 0.85 T dipole magnet, which dispersed the electron-positron pairs onto a LANEX scintillator screen, was positioned immediately after the second collimator, with the scintillator screen placed 650 mm after the rear of the dipole. The imaging system for the scintillator was placed in a light and radiation shielded box to reduce noise.

\begin{figure}[t!]
    \centering
    \includegraphics[width=0.8\textwidth]{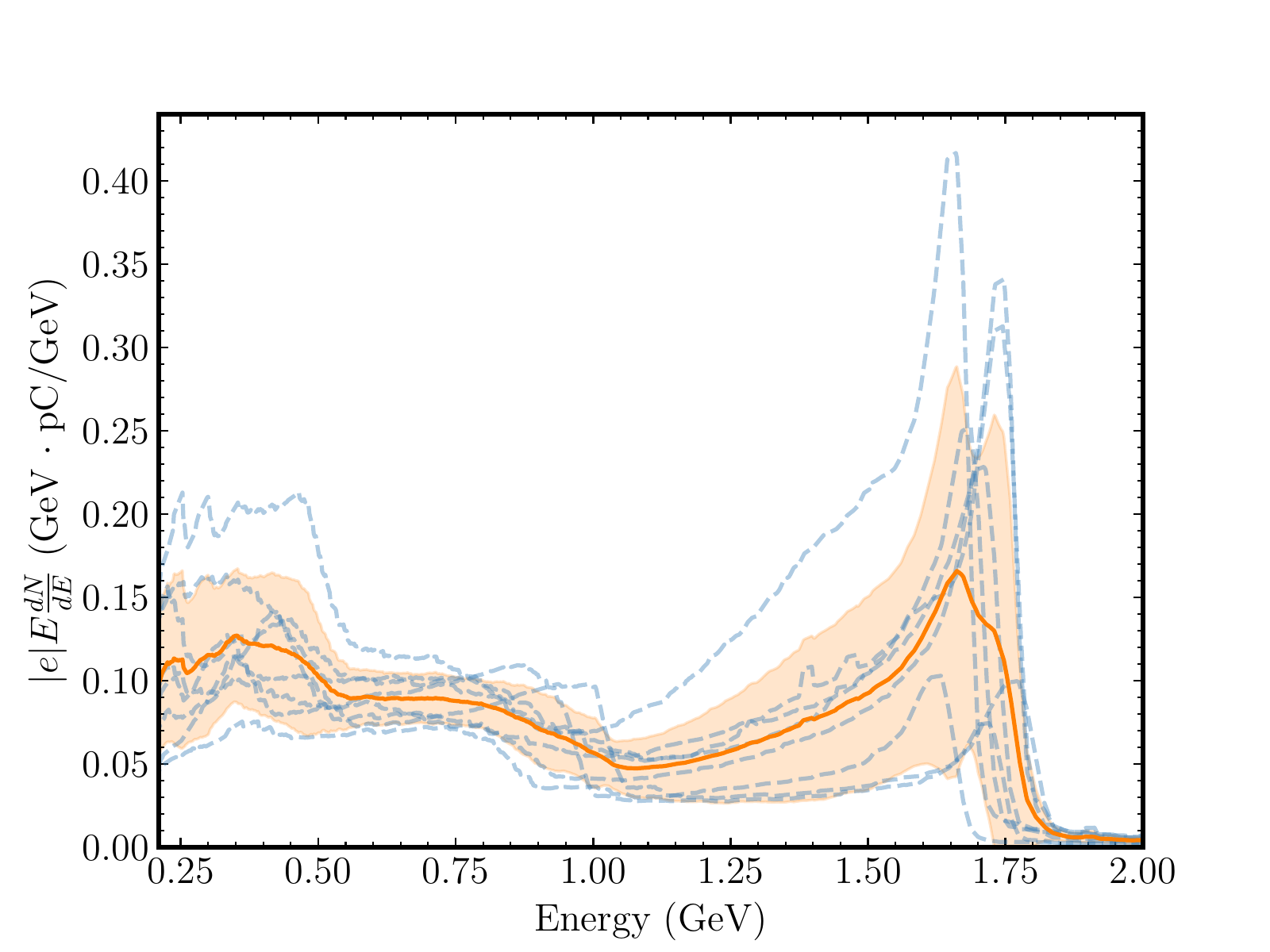}
    \caption{Examples of spectral intensity of laser wakefield electrons generated within the gas cell target for eight consecutive shots (blue dashed), their mean spectral intensity (solid orange) and one standard deviation from the mean (shaded orange).}
    \label{fig:electron_shot_spectra_IP}
\end{figure}

Typical electron spectra obtained from the laser-wakefield accelerator, after optimisation in terms of maximum charge and maximum energy, are shown in Fig. \ref{fig:electron_shot_spectra_IP}. The electron beams had a total charge above 200 MeV (lowest energy detectable by the electron spectrometer) of $207 \pm 62$ pC with a maximum  energy of $1.71 \pm 0.05$ GeV and a sub-mrad energy-dependent divergence for energies above 1 GeV ($0.5\pm0.2$ mrad at 1 GeV and $0.28\pm0.05$ mrad at 1.6 GeV). The statistical uncertainties reported here are all dominated by shot-to-shot fluctuations in the laser and plasma parameters and are taken into account in the numerical modelling reported below. These electron spectra were taken prior to any run where bremsstrahlung radiation was generated since the insertion of the converter target did not allow for a measurement of the spectrum of the electrons impinging on it. 

\begin{figure}[b!]
    \centering
    \includegraphics[width=\linewidth]{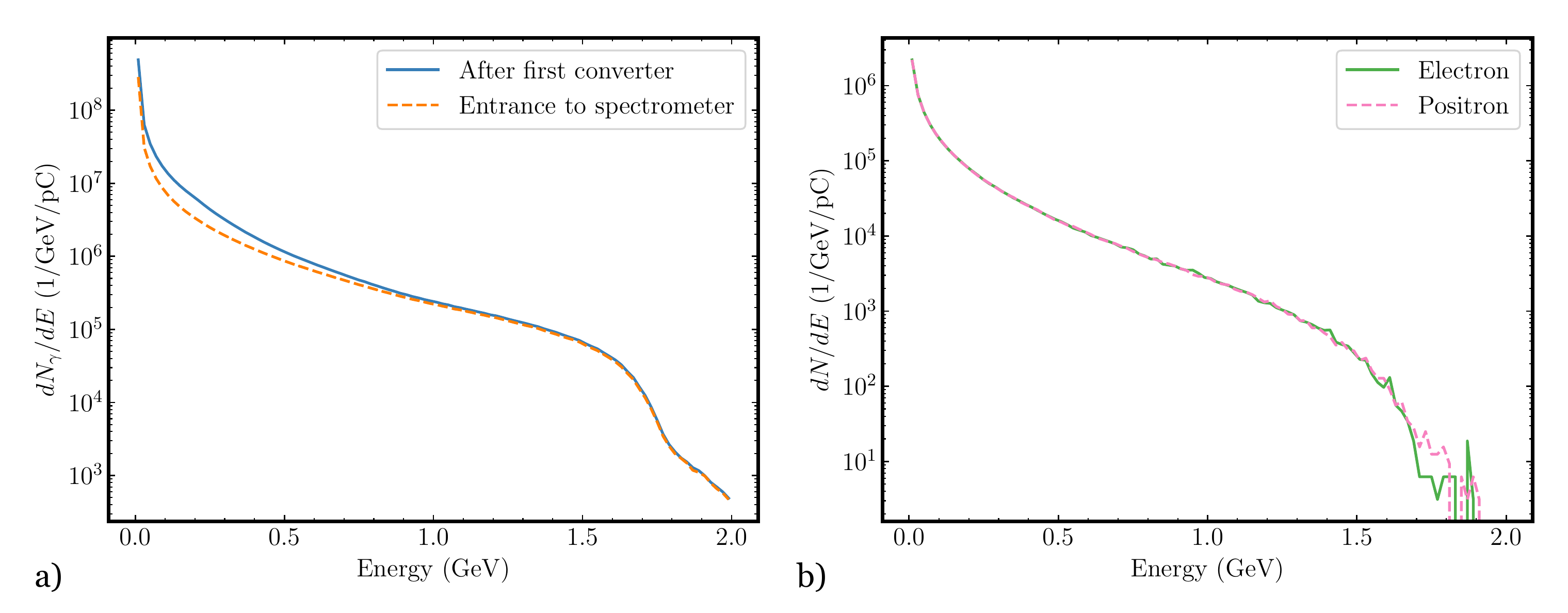}
    \caption{Simulated energy spectra per pC of electron beam charge of (a) photons exiting the converter located (blue solid), and photons incident on the spectrometer converter (orange dashed), and (b) $e^-e^+$ pairs (green solid and pink dashed, respectively) produced within the spectrometer converter which reach the LANEX detector.}
    \label{fig:g4sim_spectra}
\end{figure}

The expected spectrum of the bremsstrahlung gamma-ray beam produced during the interaction of such electron beams with a 1 mm tantalum converter was modelled using the Monte-Carlo (MC) particle tracking code Geant4 \cite{Agnostinelli:2003, Allison:2006}. The whole geometry presented in Fig. \ref{fig:Apollon_layout} along with $1\times 10^8$ primary LWFA electrons, sampled from the average electron spectrum shown in Fig. \ref{fig:electron_shot_spectra_IP}, were generated as input for the MC simulations. To show the effect of the propagation through the whole experimental setup up to the entrance of the gamma-ray spectrometer, we show in Fig. \ref{fig:g4sim_spectra} the predicted gamma-ray spectrum at the rear surface of the converter target (solid blue) and the one entering the spectrometer (dashed orange). At source, the gamma-ray photons present a monotonically decreasing spectrum up to a maximum of $\approx$1.7 GeV with a total number of photons with an energy larger than 1 MeV of $(2.21\pm 0.66) \times 10^9$ for a total charge of the electron beam of $207\pm 62$ pC (corresponding to 1.7 photons per primary electron). The gamma ray beam thus contains $22 \pm 7$ mJ of energy. The propagation through the chamber window and the air gap to the entrance of the gamma-ray spectrometer affects the photon beam only marginally, particularly at high energy. Fig. \ref{fig:g4sim_spectra} shows that for energies greater than $0.6$ GeV, the photon spectrum entering the spectrometer is virtually identical to the one at source. The total number of photons entering the spectrometer is $(1.49 \pm 0.44)\times 10^9$, with a total beam energy of $16 \pm 5$ mJ. Only for energies $\lesssim 0.6$ GeV, there is a reduction in the number of photons entering the spectrometer compared to that produced at source due to the collimation system and scattering during propagation.

For this experimental configuration, the energy resolution of the gamma-ray spectrometer is limited by the divergence of the radiation source and, hence, of the converted $e^+e^-$ pairs. Appendix \ref{Magnetic Dispersion} provides additional details; however, the energy resolution of a divergence-limited magnetic spectrometer scales as $\Delta(E)/E \propto E\Theta_S(E)/z_M B$, where $E$ is the particle energy, $\Theta_S(E)$ is the minimum between the divergence of the converted pairs and the collimator acceptance angle, and $z_M B$ is the integrated field strength of the magnet. For our configuration, this resulted in a resolution of $\Delta(E)/E = 0.26E$ [GeV]. As discussed in the following, much higher spectral resolution can be readily obtained at higher photon energies or using a stronger dipole magnet.
\section{Analysis of Results}\label{experimental_analysis}
\begin{figure}[t!]
    \centering
    \includegraphics[width=16cm]{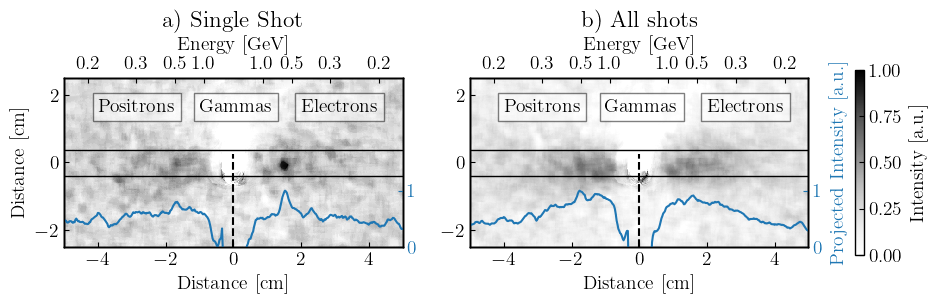}
    \caption{\textbf{a.} Example of single-shot background-subtracted data recorded by the scintillator screen in the gamma-ray spectrometer. The figure evidences the positron and electron signal region on either side of the spectrometer axis. \textbf{b.} Average over eight consecutive shots. 
    An artefact induced by the bottom edge of the collimator aperture of the spectrometer is visible in both images close to (0,0). This artefact is due to a non-ideal background subtraction around the collimator edges, due to shot-to-shot fluctuations in the pointing of the gamma beam.}
    \label{fig:rawimg}
\end{figure}

A typical example of the background-subtracted single-shot data collected on the LANEX screen at the back of the gamma-ray spectrometer is shown in Fig. \ref{fig:rawimg}.a together with an average over 8 consecutive shots in frame 4b. The data evidences a central region corresponding to the gamma-rays passing through the collimator, which have been removed by background correction, together with two bands of signal on either side of it corresponding to the dispersed positrons (left) and electrons (right) generated at the converter foil in front of the spectrometer. Due to the imaging system used, each pixel in the image corresponds to 255 $\mu$m. 

Single-shot spectra for electrons and positrons were extracted by integrating along the non-dispersion axis in the region marked by a black rectangle in Fig. \ref{fig:rawimg}(a). To account for the background, 5 consecutive shots were taken with the converter foil removed and the average was subtracted from the signal shots. 

\begin{figure}
    \centering
    \includegraphics[width=0.8\textwidth]{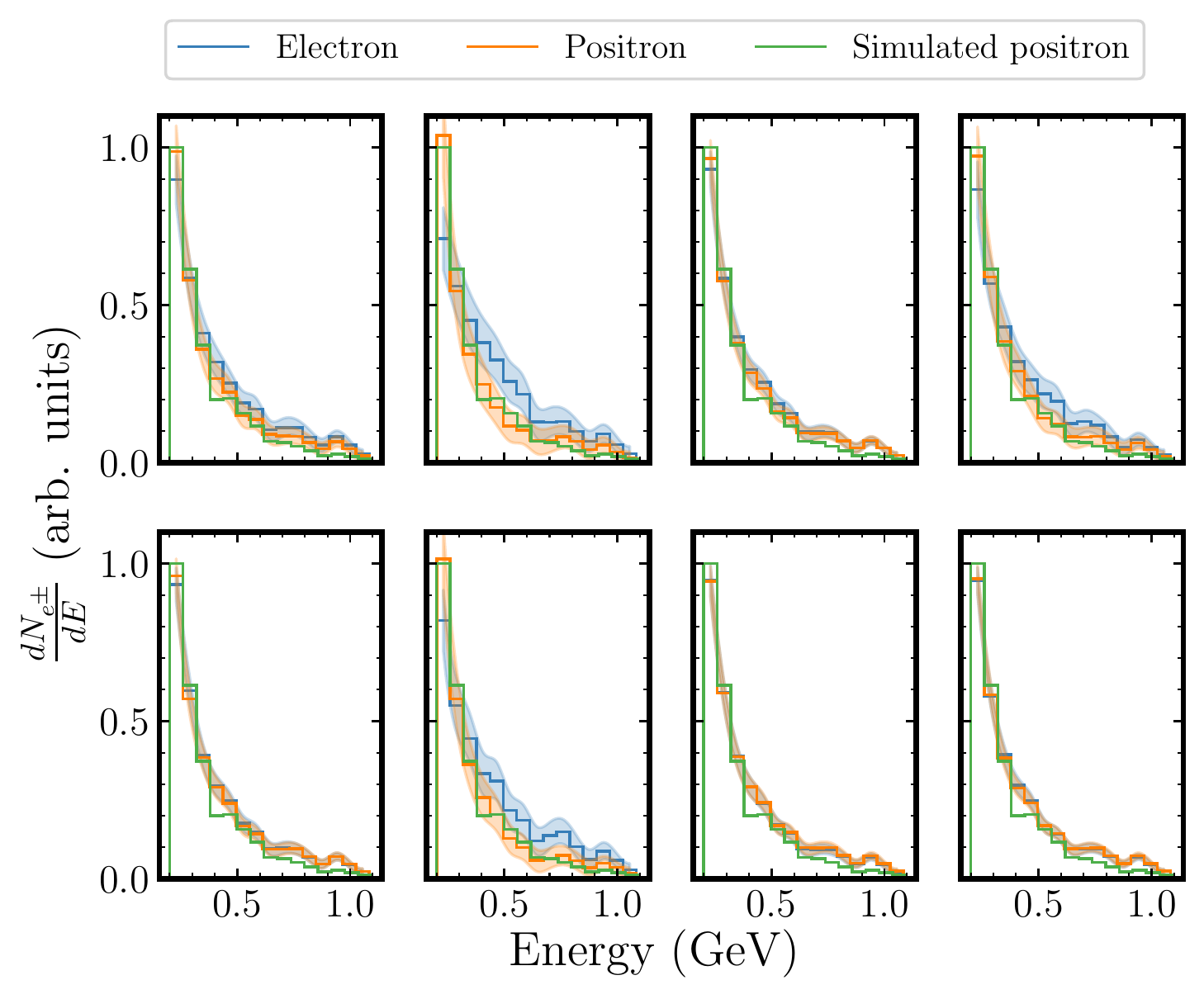}
    \caption{Energy spectrum of $e^+e^-$ pairs measured at the back of the spectrometer with corresponding uncertainty (shaded). Experimental results are compared to the positron spectrum (green) extracted from simulations.}
    \label{fig:e+e-_experimental_optimum_bins}
\end{figure}

Examples of single-shot electron (blue) and positron (orange) spectra obtained from the gamma-ray spectrometer over eight consecutive shots are shown in Fig. \ref{fig:e+e-_experimental_optimum_bins} with the shaded band corresponding to the uncertainty due to the spectral resolution of the spectrometer. For each shot, one can observe the expected behaviour of a monotonically decreasing spectrum ranging from 200 MeV (minimum energy detectable by the spectrometer) up to approximately 1.1 GeV (corresponding to the position on the LANEX screen where the signal would start to overlap with the straight-through gammas). As expected, the electron and positron spectra are indistinguishable within uncertainty. These experimental results show excellent agreement with the simulated positron spectrum at the detector plane (from Fig. \ref{fig:g4sim_spectra}), which is overplotted as a green solid line in each frame in Fig. \ref{fig:e+e-_experimental_optimum_bins}. 

The Monte Carlo simulations indicate a peak fluence on the gamma-ray spectrometer converter between $0.02$ and $0.03$ J/cm$^2$, with a mean fluence of $0.016 \pm 0.005$ J/cm$^2$. It must be noted that this is much smaller than the threshold fluence of tantalum ($\sim 0.6$ J/cm$^2$ for sustained irradiation at 30 Hz \cite{Torrisi:2000}), indicating that the spectrometer suffers negligible degradation over sustained operation. This is confirmed by our experimental findings over the course of the campaign, as no ablative effects on the converter foil were observed. This demonstrates that the gamma-ray spectrometer is capable of operating on a single-shot basis over extended periods of operation.

\section{Deconvolution of the signal}\label{deconvolution}
The measured electron and positron spectra were grouped into 15 equal bins in the energy range $[200, 1100]$ MeV for eight shots with identical LWFA interaction parameters and this data was then deconvolved using the ultra-relativistic Bethe-Heitler cross section for pair production initiated by a photon in a solid target \cite{Tsai:1974, Klein:2006}. 

The mathematical process of deconvolution involves solving an ill-conditioned inverse problem; as such, direct solution methods give results whose uncertainty is difficult to quantify, particularly in the presence of background. It is then proposed here to formulate the deconvolution process in a statistical manner, invoking the use of Bayesian statistics to not only achieve regularisation of the ill-conditioned problem, but also to allow for a measure of the reconstruction uncertainty in the form of confidence intervals. This implementation follows an iterative Bayesian method similar to d'Agostini's unfolding technique \cite{DAgnostini:2010}, and was applied to each spectrum to reconstruct the photon energy spectrum entering the spectrometer system.
Further details on the deconvolution process are given in the Appendix. While choosing to deconvolve either the electron or the positron spectrum is in principle irrelevant due to the symmetry of pair production in the converter, the electron spectrum at the detector plane is in general susceptible to a higher level of noise, due to the additional channels that can lead to electron generation or scattering (e.g., Compton scattering in the converter). We will thus focus our attention here on the deconvolution of the positron spectrum.

\begin{figure}
    \centering
    \includegraphics[width=0.85\textwidth]{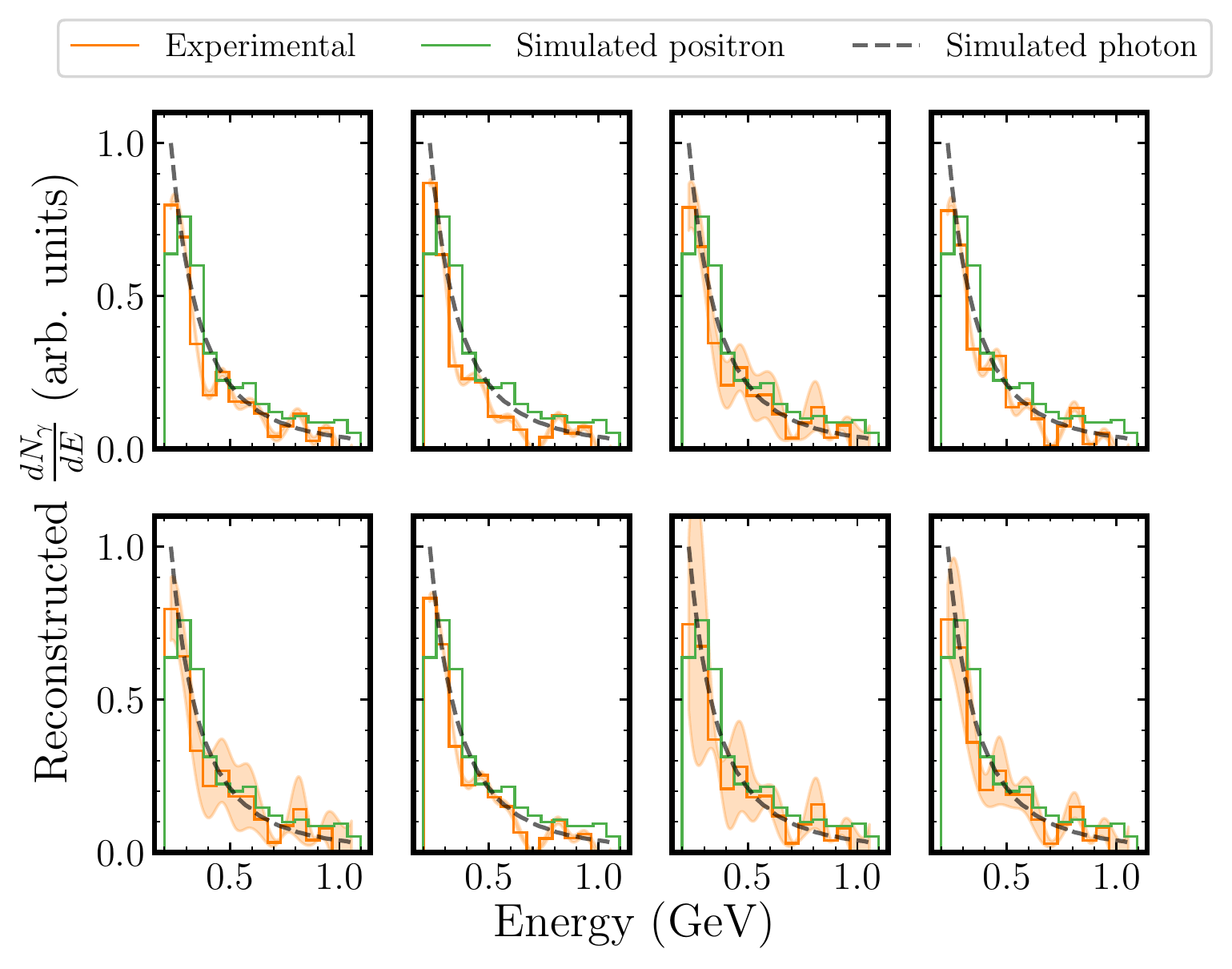}
    \caption{Reconstructed photon spectra obtained by applying the deconvolution algorithm to the experimental (orange) and simulated (green) positron spectra in Fig. \ref{fig:e+e-_experimental_optimum_bins}. The overlaid dashed line shows the photon spectrum incident on the spectrometer from simulation, as in Fig. \ref{fig:g4sim_spectra}(a). Shaded bands represent the 95\% HPDI calculated by the algorithm.}
    \label{fig:e+e-_deconvolution_optimum_bins}
\end{figure}

The results of the deconvolution process are shown in Fig. \ref{fig:e+e-_deconvolution_optimum_bins}, where the orange (blue) line represents the reconstructed gamma-ray spectrum obtained from the experimental (simulated) positron spectrum and the dashed black line is the predicted gamma-ray spectrum entering the spectrometer from Fig. \ref{fig:g4sim_spectra}(a). Although not shown, the reconstruction of the electron spectrum yields similar results and can be used as a consistency check, as described in Appendix \ref{Reconstruction Algorithm}.

 The deconvolution of the simulated positron spectrum represents the theoretical best possible reconstruction for the spectrometer setup with the same discretisation size as the simulated electron spectrum without noise. 
 The spectral shape of the incident photon spectrum is well reconstructed by this measuring technique within a 95\% confidence interval, chosen here to be the highest posterior density interval (HPDI). 
 A certain level of underestimation is consistently observed only in the lowest energy bin between 200 and 260 MeV. This can be understood by considering that there is a lack of information on the produced pairs at energies above 1.1 GeV and, therefore, limited information on the photon flux expected in this energy range. 
 The deconvolution algorithm does not consider photon energies outside of the measured range; however, these photons can still significantly contribute to pair production and this contribution is most prominent at lower energies, due to the typical spectral shape of the pair production process. 
 This observation is a further demonstration of the fact that gamma-ray spectroscopy of this kind is absolutely valid only if the full energy range of the expected photon population is resolved.  
 Nonetheless, apart from the lowest energy bin, the rest of the reconstruction shows good fidelity with the expected signal. The resulting $95\%$ HPDI of the reconstructions is a symmetric range of relative error $\sim 0.1$, increasing to $\sim 0.15$ at lower energies, another consequence of the lack of knowledge of the spectral distribution of particles with energies above $1100$ MeV.

\section{Discussion and conclusions}\label{conclusions}
While the deconvolution process is observed to introduce a small uncertainty in the reconstructed gamma-ray spectrum, the dominant contribution to the uncertainty in the gamma-ray spectrum arises from the spectral measurement of the electron-positron pairs generated in the converter.

As discussed in Section \ref{experimental_setup} and Appendix \ref{Magnetic Dispersion}, the energy resolution of the spectrometer is dominated by the divergence of the electron-positron pairs and the magnet strength: $\Delta(E)/E \propto \Theta_S(E)/(z_M B)$, with $z_M B$ the integrated field strength of the magnet.
In our experiment, the relatively low strength of the dipole magnet induced a sub-optimal energy resolution of $\Delta(E)/E = 0.26 E$[GeV].
The resolution can be increased simply by increasing the length or strength of the magnet employed. As an example, a relatively standard permanent dipole magnet of strength $1$ T and active length $150$ mm would readily give a $3.5$ factor improvement in the resolution across the 100s of MeV range. For example, this would translate to an energy uncertainty of $\sim 75$ MeV at $1$ GeV.

We also note that SFQED experiments currently being designed at large scale facilities, such as E-320 at SLAC \cite{E320} and LUXE at the Eu.XFEL \cite{Abramowicz_2021,LUXE_TDR} are anticipated to produce Compton-scattered photon beams from electron beams with an energy of 13 and 16.5 GeV, respectively. The resulting photon beam will thus present a much narrower angular distribution since this is approximately inversely proportional to the Lorentz factor of the particle at the generation point ($\theta_\gamma \sim \gamma_B^{-1}$). For example, numerical modelling of the LUXE experiment indicates that the photon beam will have an angular divergence of approximately 30 $\mu$rad, to be compared with the 510 $\mu$rad observed here. This more narrow cone and the higher energy of the photons are expected to result in an energy resolution that is 17 times lower than the one observed in this experiment, resulting in a typical energy resolution at the GeV level of the order of 1 - 2 \%.

Another crucial aspect of the spectrometer is that it must be able to operate at a single-shot level and over sustained periods of operation. Our results confirm the expectation of negligible ablation of the converter target. From our simulations, we estimate a maximum flux of the gamma-rays to be withstood by the spectrometer of $\sim 4\times10^{10}$ photons/cm$^2$/shot. This is approximately 30 times higher than the flux observed in our experiment, and would correspond to an electron beam with an overall charge of 6 nC. 

The principles of operation of the gamma-ray spectrometer reported in this article mainly follow those of the first conceptual design reported in Ref. \cite{Fleck:2020}, albeit with fundamental key differences. The most important development is in the algorithm for the reconstruction of the gamma-ray spectrum. A direct back-substitution algorithm, as originally proposed, presents fundamental issues in the regularisation of the inverse problem and results in uncertainties that are exponentially growing with each step of the algorithm. Here, we present a statistical Bayesian approach to deconvolution, which allows for a more rigorous regularisation of the inverse problem and allows one to keep lower uncertainties.  

Additionally, the experimental construction of the spectrometer required some alteration to the conceptual design due to constraints on available space and equipment, as well as the difference in performance of electron (and subsequent photon) beams generated by wakefield acceleration.

In conclusion, we report on the first high-resolution and on-shot measurement of the spectrum of high-flux GeV-scale photon beams. Our experimental results confirm, in good agreement with numerical modelling, the possibility of performing single-shot measurements of the gamma-ray spectra with a high signal to noise and $\approx$10\% energy resolution. Simple scalings with magnet strength and particle energy suggest that percent-level energy resolution can be readily achieved. We envisage that this diagnostic technique will provide key data in strong-field QED experiments such as E-320 and LUXE.

\section{Acknowledgements}
G. S. wishes to acknowledge support from EPSRC (grant numbers EP/T021659/1, EP/V044397/1, and EP/V049186/1). The authors are grateful for the support of the Apollon team and for useful discussions with members of the E-320 and LUXE collaborations.
\section{Author Contributions}\label{authors}
G.S. devised the idea and its experimental incarnation, with help from K.F., N. C., and M.J.V.S. The experimental campaign was  carried out by N.C., K.F., C.B., R.C., L.C., S.D.D., I.M., L.R., O.V. A.W., A.S., and B.C., led by M.J.V.S. and L.T.D. K.F. carried out the numerical simulations and N.C., E.G., and K.F. analysed the data. The manuscript was written by G.S., K.F., N.C., and E.G., with help from all the authors.

\bibliography{references}

\appendix
\section{Magnetic Dispersion}\label{Magnetic Dispersion}
The information obtained from the spectrometer is the position space spectrum of the electrons and positrons. The particle position $x $ can be related to its energy $E$ using the magnetic dispersion relation for a detector a distance $z_D$ from the rear of a magnet of length $z_M$ and field strength $B$ ,
\begin{equation}\label{deflection_approx}
    E(x) \simeq eB \cdot \frac{z_M}{x}\left(\frac{z_M}{2} + z_D\right),
\end{equation}
assuming $E \gg eBz_M$. Additionally, the energy resolution on the detector is limited by two factors, leading to an uncertainty in the binning process: the divergence of the original photon beam \cite{Glinec:2006}; and the spatial resolution of the detector. The total resolution available is then the quadrature sum of these components, assuming they are independent, 
\begin{equation}\label{resolution}
    \frac{\Delta(E)}{E} =  \frac{(z_S + z_M +z_D)\Theta_S}{(z_D+z_M/2)z_M} \cdot \frac{E}{eB} \oplus \frac{\delta x}{z_M(z_M + z_D)}\cdot\frac{E}{eB},
\end{equation}
where $\delta x$ is the spatial resolution of the screen (e.g. pixel width), $z_S$ is the distance between the source of the pairs (the converter foil) and the entrance of the magnet and $\Theta_S$ is the divergence of the electrons/positrons after conversion. In the ultrarelativistic limit, if the photons have divergence, $\theta_\gamma$, and the leptons have a Lorentz factor, $\gamma$, then $\Theta_S \sim \sqrt{\theta_\gamma^2 + \frac{1}{\gamma^2}}$ can be assumed.

\section{Reconstruction Algorithm}\label{Reconstruction Algorithm}
The number of $e^+e^-$ pairs generated by a photon beam passing through a thin converter target can be calculated using \cite{Tsai:1974}
 \begin{equation}\label{volterra_model}
    \frac{dN}{dE} = \frac{N_A\rho t}{A}\int\limits_E^{\omega_{max}} d\omega\, \frac{d\sigma}{dE}\frac{dN_\gamma}{d\omega}.
\end{equation}
$\omega_{max}$ is some known maximum value for the photon energy, $\frac{dN_\gamma}{d\omega}$ is the energy spectrum of the incident photon beam, $\frac{dN}{dE}$, the energy spectrum of the outgoing $e^+e^-$ pairs and $\frac{d\sigma}{dE}$ is the (energy) differential cross section for pair production. $A$ is the atomic mass, $X_0$ is the radiation length of the converter material, and $N_A$ is Avogadro's constant. $t$ is the thickness of the converter with a mass density $\rho$. In the complete screening limit, applicable to high-$Z$ materials and photon energies $\gtrsim 1$ GeV, a quadratic approximation \cite{Tsai:1974,Klein:2006} for the cross section,
\begin{equation}\label{complete_screening}
    \frac{d\sigma}{dE} = \frac{A}{N_A X_0}\frac{1}{\omega}\left[1 - \frac{4}{3}\frac{E}{\omega}\left(1-\frac{E}{\omega}\right)\right],
\end{equation}
is valid.

Then, after determination of the $e^+/e^-$ spectrum, the inversion of Eq. \eqref{volterra_model} to get the original photon spectrum can be performed by first discretising the integral
and letting $\frac{dN_\gamma}{d\omega} \rightarrow \bm{f}$ and $\frac{dN}{dE} \rightarrow \bm{g}$.
Deconvolution can then be posed as the inversion of the linear matrix equation
\begin{equation}\label{OperatorProblem}
    \bm{g} = \mathcal{K} \bm{f}.
\end{equation}
$\mathcal{K}$ is a matrix whose elements encompass the operator term
$ \frac{N_A\rho t}{A}\int d\omega \, \frac{d\sigma}{dE} \circ$; the exact structure of this kernel operator depends on the quadrature method employed. The standard trapezium rule is used in this analysis.
This problem can be solved by direct back substitution \cite{Press:2007} as the kernel is triangular ($\mathcal{K} = 0$ if $\omega < E$), however, when combined with a  Bayesian approach \cite{Piana:1994, Djafari:1993, Djafari:1996}, noise in the measured spectrum can be more robustly dealt with.

To pose the deconvolution problem in a statistical manner, a noise term, $\bm{\eta}$, is introduced to the operator equation
\begin{equation}\label{operator_problem}
    \bm{g} = \mathcal{K}\bm{f} + \bm{\eta}.
\end{equation}
A statistical distribution is then applied to the noise - this is chosen to be Gaussian and independent of the signal, i.e. $\eta_i \sim \mathcal{N}(0, \sigma_b^2)$ where $\sigma_b^2$ is the variance in the noise. From Eq. \eqref{operator_problem},
\begin{equation}
    \mathbb{P}(\bm{g}\lvert \bm{f}, \beta, \mathcal{K}) = \left(\frac{\beta}{2\pi}\right)^{\frac{n}{2}}\exp\left(-\frac{1}{2}\beta\lVert \bm{g} - \mathcal{K}\bm{f}\lVert^2\right).
\end{equation}
Here, $\mathbb{P}(g\lvert \bm{f},\beta, \mathcal{K})$ is the likelihood function of the problem with $n = \dim \bm{g} = \dim \bm{f}$ and $\beta = 1/\sigma_b^2$ is the precision. To apply Bayesian methods, a prior distribution must be stated while making only general assumptions about $\bm{f}$, namely that the values of the spectrum cannot be negative, i.e. $\bm{f} \succeq 0$. Hence, a normal distribution with unknown variance to be inferred, and mean determined by the back substitution solution to \eqref{OperatorProblem}. By construction, this prior mean will be strictly non-negative in all components.

An application of empirical Bayes inference which involves maximising the marginal likelihood function to infer the unknown variances (equivalently, precisions),
\begin{equation}
    \mathbb{P}(\bm{g}\lvert \beta, \theta, \bm{f}_0, \mathcal{K}) = \int d^nf \, \mathbb{P}(g\lvert \bm{f},\beta, \mathcal{K}) \cdot \mathbb{P}(\bm{f}\lvert \bm{f}_0, \theta),
\end{equation}
can then be used. These inferred values are then used to calculate the resulting posterior distribution, $\mathbb{P}(\bm{f}\lvert \bm{g}, \beta, \theta, \bm{f}_0, \mathcal{K})$. The mean (maximum a posteriori estimate) of this distribution is taken to be the solution to the inverse problem \eqref{operator_problem} while the error associated with the reconstruction of the spectrum is represented by the highest posterior density interval (HPDI) about the solution. 

\begin{figure}
    \centering
    \includegraphics[width=\textwidth]{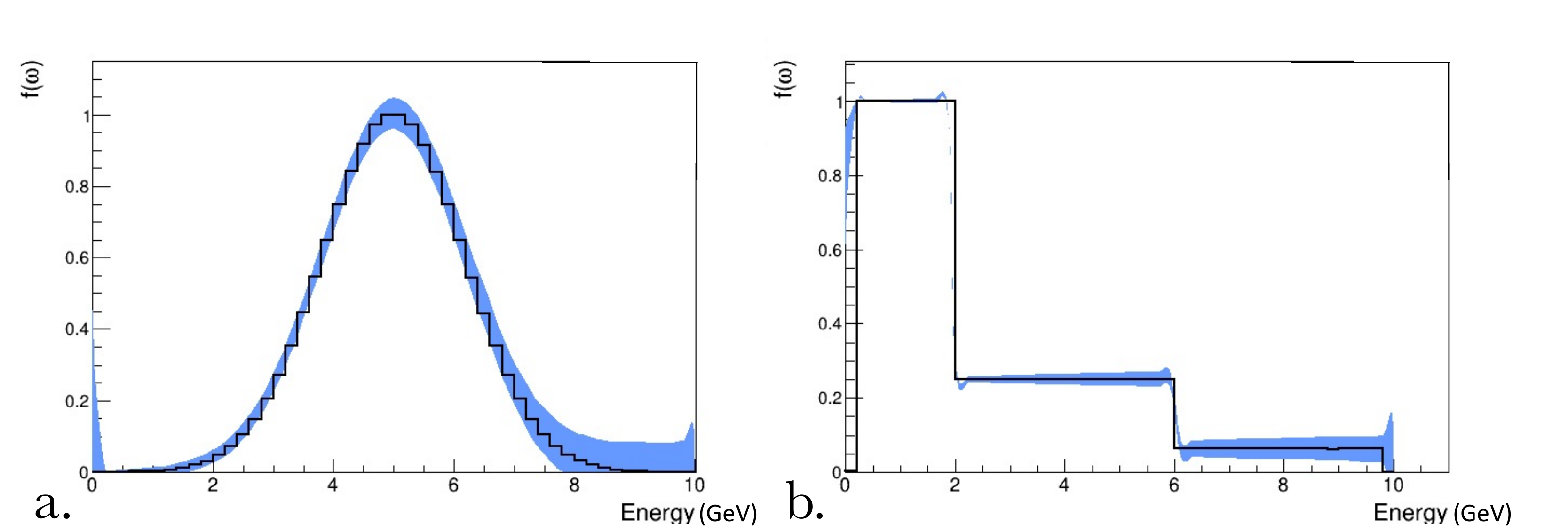}
    \caption{Result of deconvolving (a) a Gaussian (mean = $5$ GeV, $\sigma = 1.2$ GeV), and (b) a piecewise step-like photon spectrum. Black is the original spectrum and the shaded blue is the 95\% HPDI of the reconstruction in each case.}
    \label{fig:algorithm_test}
\end{figure}

An example of the performance of the deconvolution algorithm is given in Fig. \ref{fig:algorithm_test}. In these tests, a Gaussian (frame a.) and a piecewise step function (frame b.) were used as photon spectra, which were then propagated through the spectrometer in simulation. The resulting electron and positron spectra were then deconvolved using the above approach, to give the retrieved gamma-ray spectrum, with associated uncertainties (blue bands in Fig. \ref{fig:algorithm_test}). As can be seen from these two examples, the spectrometer is able to reconstruct the gamma-ray spectrum with a high fidelity, and able to precisely identify the edges in the spectrum. For example, the edge at 2 GeV in Fig. \ref{fig:algorithm_test}.b is identified with an uncertainty of $\pm$ 120 MeV ($\approx$ 6\%), while the uncertainty in identifying the edge at 6 GeV is of the order of $\pm$ 170 MeV ($\approx$ 3\%).

\end{document}